\documentclass[prb,twocolumn,superscriptaddress]{revtex4-1}

\usepackage{amsmath}
\usepackage{amssymb}
\usepackage{xspace}
\usepackage{graphicx}
\usepackage{grffile}
\usepackage{nicefrac}
\usepackage{color}

\graphicspath{{figs/}}

\begin{document}

\title{Interplay of charge-transfer and Mott-Hubbard physics approached
by an efficient combination of self-interaction correction and dynamical mean-field theory}

\author{Frank Lechermann}
\affiliation{I. Institut f{\"u}r Theoretische Physik, Universit{\"a}t Hamburg, 
Jungiusstr. 9, 20355 Hamburg, Germany}
\author{Wolfgang K\"orner}
\affiliation{Fraunhofer-Institut f\"ur Werkstoffechanik IWM, 
W\"ohlerstr. 11, 79108 Freiburg, Germany}
\author{Daniel F. Urban}
\affiliation{Fraunhofer-Institut f\"ur Werkstoffechanik IWM, 
W\"ohlerstr. 11, 79108 Freiburg, Germany}
\author{Christian Els\"asser}
\affiliation{Fraunhofer-Institut f\"ur Werkstoffechanik IWM, 
W\"ohlerstr. 11, 79108 Freiburg, Germany}
\affiliation{Universit\"at Freiburg, Freiburger Materialforschungszentrum FMF, 
Stefan-Meier-Str. 21, 79104 Freiburg, Germany}

\begin{abstract}
Late transition-metal oxides with small charge-transfer energy $\Delta$ raise issues
for state-of-the-art correlated electronic structure schemes such as 
the combination of density functional theory (DFT) with dynamical mean-field theory (DMFT). 
The accentuated role of the oxygen valence orbitals in these compounds asks for an enhanced 
description of ligand-based correlations. Utilizing the rocksalt-like NiO as an example, we 
present an advancement of charge self-consistent DFT+DMFT by including self-interaction 
correction (SIC) applied to oxygen. This introduces explicit onsite O correlations as well 
as an improved treatment of intersite $p-d$ correlations. Due to the 
efficient SIC incorporation in a pseudopotential form, the DFT+sicDMFT framework is an 
advanced but still versatile method to address the interplay of charge-transfer and 
Mott-Hubbard physics. We revisit the spectral features of stoichiometric NiO and reveal
the qualitative sufficiency of local DMFT self-energies in describing spectral 
peak structures usually associated with explicit nonlocal processes. For Li$_x$Ni$_{1-x}$O, 
prominent in-gap states are verified by the present theoretical study.
\end{abstract}

\maketitle

\section{Introduction}
An understanding of the electronic structure and phase diagrams of transition-metal (TM) 
oxides still poses a major challenge. The realistic modeling of 
strong electronic correlations in this family of compounds has been tremendously improved 
since establishing the combination of density functional theory (DFT) with dynamical 
mean-field theory (DMFT). Especially early-TM oxides with a partially-filled(empty) 
$t_{2g}$($e_g$) sub-manifold of the TM-$d$ shell, showing dominant Mott-Hubbard physics,
often may readily be addressed by such an approach. Late-TM oxides display 
a more intriguing interplay of hybridization and correlation effects and remain demanding.
Reason is that in the latter systems with usually completely-filled $t_{2g}$ subshell and 
partially-filled $e_g$ subshell, the ligand oxygen $2p$ states are much closer in energy to 
the TM $d$ states. The charge-transfer energy $\Delta$, required to transfer a ligand 
electron to a TM $d$ orbital, eventually becomes smaller than the energy cost for a
doubly-occupied $d$ orbital, i.e. the Hubbard $U$. In late TMs, the $d$ states are 
lower in energy, hence $\Delta$ shrinks, but $U$ on the other hand increases due the 
reduced orbital extension with larger nuclear charge~\cite{ima98}. According to the famous 
Zaanen-Sawatzky-Allen (ZSA) scheme~\cite{zaa85}, correlation-induced insulators with 
$\Delta<U$ are of charge-transfer kind.

While e.g. Cu oxides are typical charge-transfer compounds, the rocksalt-like NiO, a material 
of paramount importance for the development of quantum solid-state 
theory~\cite{boe37,*mot37}, is located in the intermediate regime of the ZSA diagram. 
This renders NiO particularly interesting, and more generally, marks nickelate compounds as 
exceedingly affected by various competing instabilities (see e.g. Ref.~\onlinecite{ima98} 
for a review). Stoichiometric nickel oxide is insulating with a sizeable charge gap 
of $\sim$4.3\,eV~\cite{new59,pow70,saw84,huf84} and becomes antiferromagnetic below 
$T_{\rm N}=523$\,K~\cite{rot58}. Substitutional doping of Li$^+$ for Ni$^{2+}$ is 
effective in providing holes to cause 
conductivity~\cite{hei57,goo58,kui89,elp92,rei95,lan07,zha18}.

Numerous theoretical studies examined the interacting electronic structure of NiO.  
Advanced descriptions need to rely on a sophisticated treatment of $\Delta$, $U$ and further 
key quantum characteristics of the compound. In addition to cluster 
calculations~\cite{fuj84,elp92,vee93,tag08,kuo17}, 
DFT+DMFT~\cite{ren06,kun07,yin08,miu08,nek12,thu12,leo16,pan16} and 
variational-cluster-approximation~\cite{eder05} studies 
provided already a good account of the electronic spectrum, but challenges 
remain~\cite{kuo17}. For instance, to a certain degree, the correct determination and 
understanding of the specific nature/positioning of peaks within the valence-band spectrum 
is still a matter of debate. Furthermore, the increased importance of the role of 
ligand orbitals and their hybridization with TM orbitals in materials which lack a pure 
Mott-Hubbard character, renders an investigation of defect properties very demanding.

Therefore, the intention of this work is twofold. Using NiO as a test case, we first show
that an improved description of the intriguing interplay between Mott-Hubbard and 
charge-transfer physics at stoichiometry is achieved by treating electronic 
correlations on the TM {\sl as well as} on the ligand sites. This is realized by 
an efficient combination of the self-interaction correction (SIC) 
with the charge self-consistent DFT+DMFT framework. Thereby, SIC is applied on O and
Ni marks the DMFT impurity problem. This theory advancement enables us to shed novel 
light on the intriguing features of the paramagnetic NiO spectrum. Second, the usefulness 
of this DFT+sicDMFT scheme for advanced correlated materials science is demonstrated by 
the application to the even more challenging case of Li-doped NiO. We straightforwardly 
reproduce the long-standing experimental finding of in-gap states at $\sim 1.2$\,eV 
above the valence-band maximum.

\section{Theoretical Approach}
\subsection{Problem and general framework}
In simplest approximation~\cite{ima98}, the charge-transfer energy is given by the 
difference between the TM$(d)$ and O$(2p)$ single-particle levels, i.e. 
$\Delta=\varepsilon_d-\varepsilon_p$. In early TM oxides, such as e.g. certain titanates 
or vanadates, the respective level separation is rather large with TM$(d)$ well above
O$(2p)$. Then usually, electronic correlations matter most within the partially filled
$t_{2g}$ subshell of TM$(d)$, since O$(2p)$ states are much deeper in energy than the
scale of the lower-Hubbard-band formation. The local Hubbard $U_{dd}:=U$ within TM$(d)$ 
is the smaller energy and thus governs the correlation effects. On the contrary for late TM 
oxides, $\Delta$ becomes the smaller energy and O$(2p)$ are often located {\sl between} the 
lower- and upper-Hubbard-band formation scales. 

In the latter case, the correlation physics  is more subtle. 
Of course, $U$ remains a vital player, since it triggers strong correlation. However, the
charge-transfer energy and more generally also the Coulomb interactions of local O$(2p)$ 
kind and of intersite TM($d$)-O$(2p)$ kind gain significant impact. Extended
model Hamiltonians, besides $U_{dd}$ terms furthermore including additional onsite $U_{pp}$ 
terms on oxygen and intersite $U_{pd}$ terms, are believed to become relevant for a generic
description of the correlated electronic structure~\cite{han11,han14,pan17}. But especially in
a realistic context, as e.g. DFT+DMFT, such extensions will raise issues: at least two 
new Coulomb parameters have to be quantified and questions concerning the quality of the
many-body treatment adequate for the new terms arise. Moreover, the technical/numerical 
effort increases significantly, particularly for low-symmetry structures, 
heterostructure problems, defect properties, etc..

Therefore, we here introduce an efficient extension to the state-of-the-art 
charge self-consistent DFT+DMFT framework, geared to treat charge-transfer physics with a 
minimum of additional Coulomb parametrization and essentially without any further increase 
in technical/numerical effort. First, our DFT and DMFT parts remain structurally unmodified, 
i.e. a mixed-basis pseudopotential framework~\cite{els90,lec02,mbpp_code} is utilized for the 
former and a continuous-time quantum Monte Carlo technique~\cite{rub05,wer06} for the 
latter (see Ref.~\onlinecite{gri12} for further details). Besides these established 
building blocks, a third one is intergrated. A self-interaction-correction (SIC) 
formalism~\cite{per81,fil03} is applied to cope with the correlations explicitly originating 
from the ligand sites. Importantly, the SIC procedure is performed on the pseudopotential 
level~\cite{vog96,kor10} and enters the calculational scheme simply as a modified 
ligand pseudopotential. Hence during the self-consistency cycle, the SIC effects on the 
crystal lattice adapt to the system-dependent characteristics. Those effects, without the 
coupling to DMFT here described on the effective single-particle level, capture both, the 
impact of the ligand-onsite $U_{pp}$ and of the ligand-TM-intersite $U_{pd}$. Yet effectively,
only a single additional parameter, namely the degree of SIC renormalization for O$(2p)$, 
may be sufficient.
Last but not least, DFT+sicDMFT allows to use the identical double-counting approach
as for standard DFT+DMFT: the SIC formalism is double-counting free by definition, and the
remaining TM-onsite double-counting representation remains unaffected. This is favorable
as the issue of double counting for charge-transfer systems has been a matter of 
debate~\cite{kar10,sak13}.

\subsection{Calculational details}
For the DFT part, the local density approximation (LDA) is utilized. Norm-conserving 
pseudopotentials according to Vanderbilt~\cite{van85} are constructed for Ni, O and
Li. The mixed basis consists of localized orbitals for Ni($3d$), O($2s$) and O($2p$),
as well as plane waves with an energy cutoff $E_{\rm cut}$. The self-interaction correction
for the O pseudopotential employs orbital weight factors $w({\rm O})=(w_s,w_p,w_d)$
and a global SIC scaling parameter $\alpha$ (see Refs.~\onlinecite{vog96,kor10} for 
further details). 
While by default the O($2s$) potential is 100\% corrected and the 
O($3d$) potential is 0\% corrected, key freedom is provided by the correction $w_p$ of the 
crucial O($2p$) potential, i.e. $w({\rm O})=(1.0,w_p,0.0)$.
In the case of an isolated atom $\alpha=1$ (atomic SIC) holds. A detailed 
study of the variation of $\alpha$ in different kind of solids is given by 
Pemmaraju et al.~\cite{pem07}.
Throughout the work we choose $\alpha=0.8$, shown to be reliable in previous calculations 
for TM oxides~\cite{kor10,kor14,urb16}. If not otherwise stated, we also pick $w_p=\alpha$
and thus effectively, a single additional parameter setting enters here
the advanced formalism. 
Note that no self-interaction correction is applied to Ni, 
since the whole localization/correlation effects originating from the nickel site are 
here described within DMFT.

Projected local orbitals~\cite{ama08,ani05,hau10} are employed to define 
the correlated subspace for the DMFT part. The five Ni($3d$) atomic-like orbitals are
projected onto eight Kohn-Sham valence states of NiO, namely onto the dominant Ni$(3d)$ and 
O$(2p)$ dispersions. Remaining states are of course included in the complete charge
self-consistent framework, but do not enter explicitly the correlated subspace.
The general Slater-Condon Hamiltonian 
\begin{align}
\mathcal{H}=\frac{1}{2}\sum_{\substack{m_1 m_2 \\m_3 m_4}}\sum_{\sigma\sigma'} 
U_{m_1 m_2 m_3 m_4} c^\dagger_{m_1\sigma} c^\dagger_{m_2\sigma{'}}
c^{\hfill}_{m_4\sigma{'}} c^{\hfill}_{m_3\sigma}\;,
\label{eqn:slatco}
\end{align}
with $m_i=1\ldots5$ and $\sigma,\sigma'$=$\uparrow$,$\downarrow$, is used for the 
electron-electron interaction in the correlated subspace.
Coulomb matrix elements for $l=2$ are expressed in spherical symmetry via standard Slater 
integrals $F^{k}$ through
\begin{align}
U_{m_1 m_2 m_3 m_4}   = \sum_{k=0}^{2l} a_k(m_1,m_2,m_3,m_4)\,F^k\quad,
  \label{eqn:SC-parametrisation}
\end{align}
with expansion coefficients $a_k$ given by
\begin{align}
  a_k(m_1,m_2,m_3,m_4) =& \sum_{q=-k}^k (2l+1)^2 (-1)^{m_1+q+m_2}\nonumber \\ 
  &\hspace*{-3.5cm}\times
  \begin{pmatrix}
    l & k & l \\
    0 & 0 & 0
  \end{pmatrix}^2
  \begin{pmatrix}
    l & k & l \\
    -m_1 & q & m_3
  \end{pmatrix}
  \begin{pmatrix}
    l & k & l \\
    -m_2 & -q & m_4
  \end{pmatrix}\;,
\end{align}
and parametrized using the Hubbard $U$ and Hund's exchange $J_{\rm H}$ via
\begin{equation}
F^{0}=U\;,\;\,F^{2}=\frac{14}{1+r}J_{\rm H}\;,\;\,F^{4}=rF^{2}\quad.
\end{equation}
The $F^{4}/F^{2}$ Slater-integral ratio is chosen as $r=0.625$, which is adequate for 
transition-metal atoms. For the present NiO study, $U=10$\,eV and $J_{\rm H}=1.0$\,eV are 
employed. While the latter value is standard for this compound, the Hubbard
interaction is somewhat larger than the usual value of $\sim$8\,eV~\cite{ren06,kun07}. 
But charge self-consistent
DFT+DMFT often enforces an enhanced local Coulomb interaction compared to one-shot
calculations~\cite{lec18} because of the increased number of screening channels. A recent
computation~\cite{pan17} of NiO Coulomb parameters for a $dp$ Hamiltonian within the 
constrained random-phase approximation yields also a Hubbard $U\sim 10$\,eV.  
The DMFT problem is solved by hybridization-expansion continuous-time quantum 
Monte Carlo~\cite{wer06}, as implemented in the TRIQS package~\cite{par15,set16}. A 
double-counting correction of fully-localized-limit (FLL) type~\cite{ani93} is applied. 
All calculations are performed in the paramagnetic regime at a system temperature of 
$T=580$\,K. In order to obtain the spectral function $A(\omega)=-1/\pi\,{\rm Im}\,G(\omega)$, 
analytical continuation of the Green's function $G$ from the Matsubara axis to the 
real-frequency axis is performed by the maximum-entropy method. To reveal {\bf k}-resolved spectra,
we use the Pad{\'e} method applied to the Ni self-energy for the analytical continuation. If
not otherwise stated, the shown spectra is based on the maximum-entropy method.

The NiO lattice constant is set to the experimental value~\cite{bre25} $a=4.17$\,\AA. 
Doping with lithium is realized by means of a supercell approach. Each symmetry-inequivalent Ni 
site poses a different single-site DMFT problem; all are coupled within the general 
multi-site many-body scheme~\cite{pot99}. As DFT convergence parameters we used 
$E_{\rm cut}=16(13)$\,Ryd and a {\bf k}-point mesh of $13(5)\times13(5)\times13(5)$ for the 
pristine(Li-doped) case.

\section{Results}

\subsection{Self-interaction corrected oxygen in NiO}
Before dvelving into the results of the new DFT+sicDMFT scheme, it is illustrative
to inspect the impact of the self-interaction correction to oxygen in nonmagnetic NiO from 
the effective single-particle LDA+SIC viewpoint. 

\begin{figure}[t]
\includegraphics*[width=8.25cm]{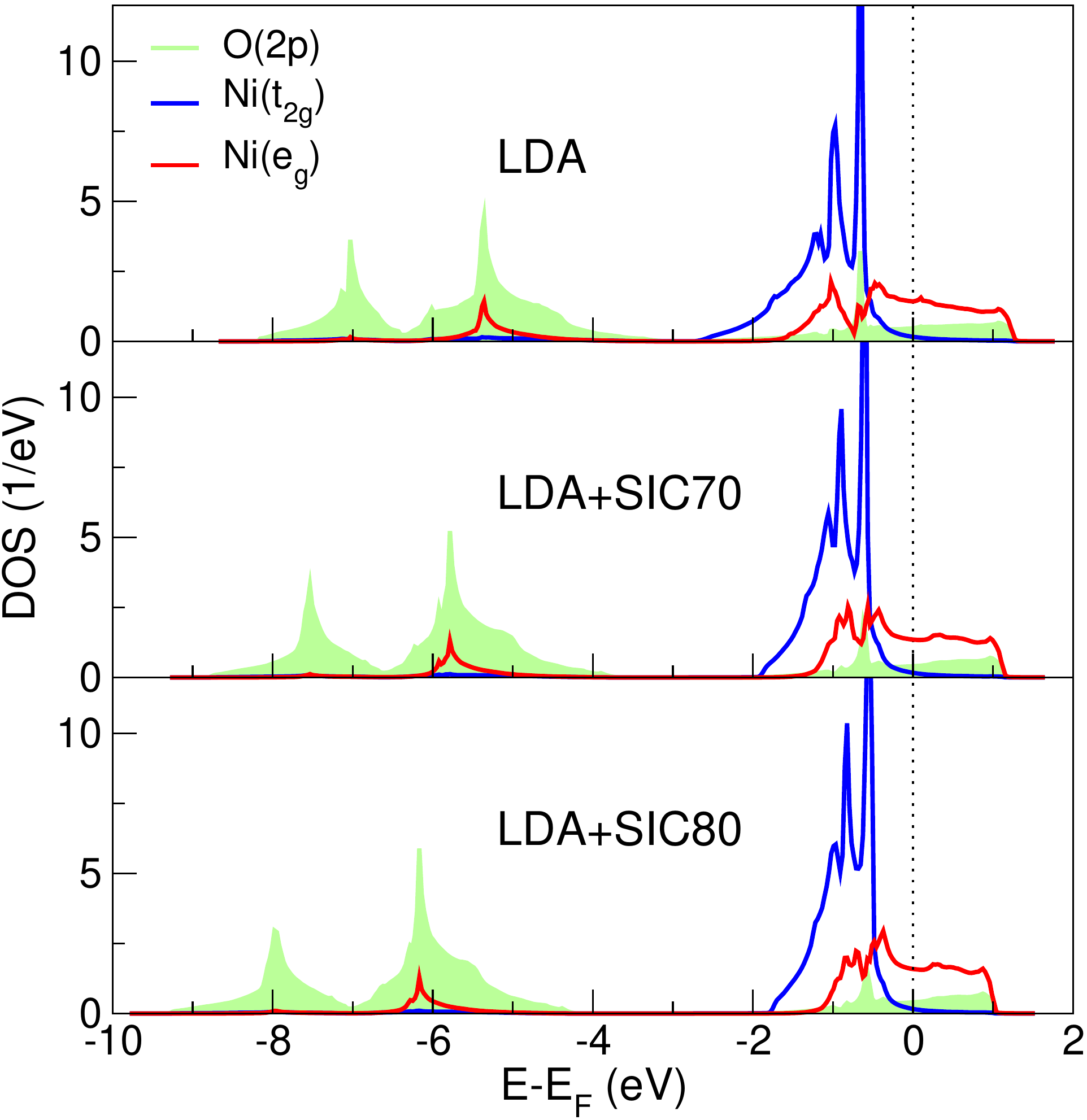}
\caption{(color online) Site- and orbital-projected DOS of NiO within LDA
(top), LDA+SIC for $w_p=0.7$ (middle) and for $w_p=0.8$ (bottom).}
\label{fig:sic}
\end{figure}

Figure~\ref{fig:sic} displays the site- and orbital-projected density of states
(DOS) within LDA and for two different SIC parameters $w_p=0.7,0.8$. The Ni-$t_{2g}$
states are completely filled and the strongly hybridized $\{$Ni-$e_{g}$,O($2p$)$\}$ states 
are partially filled. Hence holes are located on the TM site as well as on the ligand sites.
Not surprisingly, each scheme renders the system metallic. It is well known that for gap 
opening on the static DFT(+U) level, symmetry breaking in the form of magnetic ordering is 
indispensable and the same applies here. Note that we could apply SIC also on the TM 
site~\cite{kor10,urb16} and additionally allow for antiferromagnetic order to investigate 
the insulating state in that effective-single-particle approximation. The result would 
be qualitatively similar to the one obtained from a DFT+U treatment~\cite{ani91_2}, but 
we do not want to follow this route in the present work.

Instead, let us concentrate on the principal effects of SIC applied on the oxygen site. There 
are in essence two main effects. First, the $pd$ splitting and hence the charge-transfer
energy is increased by SIC. Using $\Delta=\varepsilon_d-\varepsilon_p$, the value
reads, respectively, $\Delta_{\rm LDA}=3.13$\,eV, $\Delta_{w_p=0.7}=3.98$\,eV
and $\Delta_{w_p=0.8}=4.50$\,eV. Fits to experimental data yield $\Delta$ values 
in the range $\sim[4,5]$\,eV~\cite{fuj84,laa86,kuo17}. Thus LDA severely underestimates
$\Delta$ and the effect of SIC on oxygen brings the charge-transfer energy in line with 
experimental estimates. Second, band narrowing takes place with SIC, roughly on the order of 
$Z\sim 0.8$ for $w_p=0.8$, as also visible from Fig.~\ref{fig:sic}. 
Importantly, the band-narrowing effects are not only encountered for the O($2p$) contribution, 
but also for the dominant Ni($3d$) bands. Hence a nonlocal ligand-TM correlation 
effect occurs as a result of self-interaction correction applied to the oxygen 
pseudopotential.
Those two key effects originate from the effective inclusion of $U_{pp}$ and $U_{pd}$
terms within the present LDA+SIC treatment.

Let us remark again that the parameter setting $\alpha=0.8$ is not specifically adjusted
to NiO, but this $\alpha$ value turns out to be a proper choice for various TM 
oxides~\cite{kor10,kor14,urb16}. 
In other words, a present SIC parametrization with $w_p=\alpha$ is much less case 
sensitive as the usual choice/calculation of the Hubbard interaction(s).

\subsection{NiO many-body spectrum}
\subsubsection{DFT+sicDMFT examination}
\begin{figure}[t]
\includegraphics*[width=8.5cm]{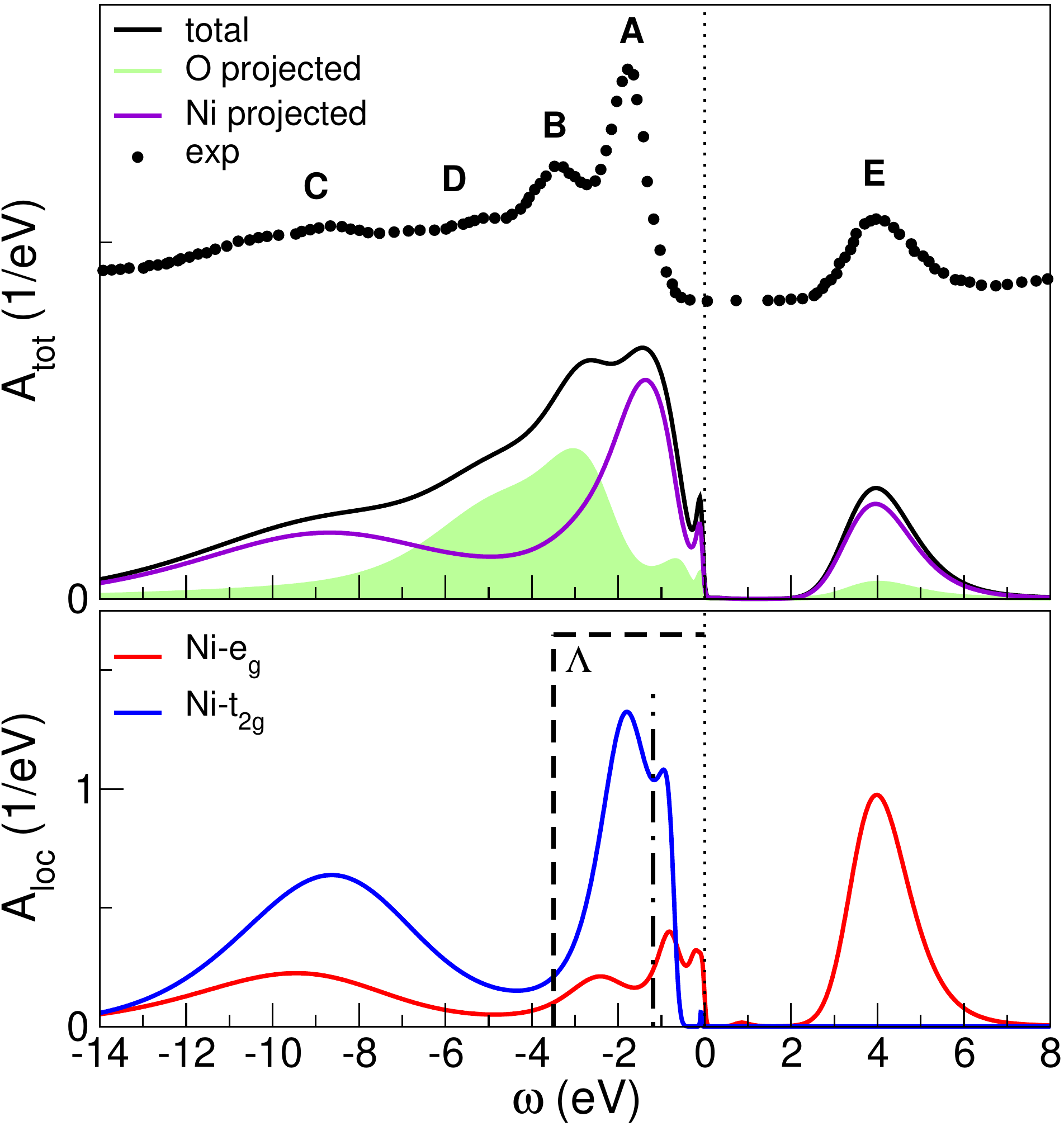}
\caption{(color online) NiO {\bf k}-integrated spectral function from DFT+sicDMFT. 
Top: total spectrum with site and orbital projection as well as comparison to 
photoemission and inverse photoemission data~\cite{saw84} at a photon energy of 1486.6\,eV. 
Points A$-$E mark specific spectral features (see text). Bottom: local Ni($3d$) 
spectrum with $e_g$ and $t_{2g}$ character. Dashed-line rectangle represents energy 
window $\Lambda$ with competing $\{d^8\underline{Z},d^8\underline{L}\}$ states (see text). 
Dot-dashed line denotes the $d^8\underline{Z}$-dominated region.}
\label{fig:totspec}
\end{figure}

We now discuss the spectral results from the complete DFT+sicDMFT approach, employing 
$w_p=0.8$. Figure~\ref{fig:totspec} shows the main outcome for stoichiometric NiO
together with the combined experimental data~\cite{saw84} from photoemission and 
inverse-photoemission.

Let us first briefly recall the state-of-the-art interpretation~\cite{fuj84,saw84,vee93} of 
the experimental spectrum. NiO is a correlation-induced insulator subject to the 
interplay of Mott-Hubbard and charge-transfer physics. The Ni ground state in the
$3d$-shell amounts to $d^8$, and $\underline{L}$ describes a hole in the ligand O($2p$) 
states. The crucial charge-transfer process, associated with the energy $\Delta$, is 
described by the transition 
$d^8\,\rightarrow\,d^{9}\underline{L}$, i.e. electron transfer from O($2p$) to Ni($3d$). 
This energy scale sets the NiO charge gap of $\sim 4.3$\,eV. An added electron conclusively
enters the $d^9$ state associated with peak E. On the other hand, adding a hole to the system
either results in the high-energy $d^7$ state (i.e. lower Hubbard band) associated with
peak C, or gives rise to $d^{8}\underline{L}$ at a much lower energy of peak A. The shoulder
D is usually\cite{saw84} interpreted as originating from the nonbonding part of O($2p$).
The most-controversially discussed peak B is build from a substructure of the 
$d^{8}\underline{L}$ state and often associated with nonlocal excitations~\cite{vee93}
(see Ref.~\onlinecite{kuo17} for a detailed discussion). 
Let us note that 
Taguchi {\sl et al.}~\cite{tag08} suggested an alternative scenario for the lower-energy 
peaks A and B. Namely, peak A should originate from a $d^8\underline{Z}$ state and only peak B
builds up from $d^8\underline{L}$, without the need of invoking explicit nonlocal
processes. Here, $d^8\underline{Z}$ refers to a Zhang-Rice (ZR) bound state, a strongly 
correlated TM-O-hybridized low-energy entity, presumably most relevant for low-energy 
cuprate physics~\cite{zha88}. In NiO, the ZR (doublet) bound state is based on the interaction 
of the O($2p$) hole with both Ni($3d$) holes. Loosely speaking, $d^8\underline{Z}$ is the 
tightly-bound 'collective' counterpart of the weakly-bound $d^8\underline{L}$ excitation.

The upper part of Fig.~\ref{fig:totspec} displays the total {\bf k}-integrated spectral function 
$A_{\rm tot}(\omega)=\sum_\nu A_{\nu}(\omega)$, with $\nu$ as the Bloch (band) index, 
from DFT+sicDMFT compared to the experimental data. Additionally, the projection of 
$A_{\rm tot}(\omega)$ onto Ni($3d$) and O($2p$) is depicted. Note that this site- and 
orbital-resolved spectrum is strictly not identical to the true local spectral function
in a many-body sense. Since the projection is performed from the Bloch-resolved 
$A_{\nu}(\omega)$ it carries the complete hybridization on the lattice and moreover 
results from an analytical continuation of the Bloch Green's function $G_{\nu}$. Overall, 
the agreement with experiment is quite remarkable: the theoretical charge gap of $\sim 4$\,eV 
matches perfectly, and also the further features A$-$D are well reproduced. The principal 
charge-transfer character is obvious from the fact that the dominant part of O($2p$) is 
located between the lower Hubbard band at $\sim -9$\,eV and the upper Hubbard band at 
$\sim 4$\,eV. As already expected from the previous discussion, the lower-energy region 
$\Lambda=[-3.5,0]$\,eV indeed asks for a deeper analysis. 

Therefore, the lower part of Fig.~\ref{fig:totspec} shows the Ni($3d$) local spectral 
function $A_{\rm loc}(\omega)$ as obtained from analytical continuation of the local Green's 
function $G_m$. Locally, the $t_{2g}$ manifold is completely filled and the upper Hubbard 
band is exclusively of $e_g$ character. Within $\Lambda$ the $e_g$ part displays a 
three-peak structure, whereas the $t_{2g}$ part a two-peak structure. The first sharp 
resonance closest to the valence-band maximum (VBM) at $\sim-0.1$\,eV is of exclusive $e_g$ 
kind, in line with previous studies~\cite{elp92,kun07,yin08,kuo17}. A second sharp peak 
resonates in both cubic $3d$ sectors roughly at the same energy $\sim-0.85$\,eV, while 
a slightly broader peak occurs at $\sim-1.8(-2.5)$\,eV for $t_{2g}$($e_g$). Albeit a bit 
speculatively, we interpret this intriguing structure as follows: 
the first sharp substructure at $\ge -1.2$\,eV belongs to 
$d^8\underline{Z}$, whereas the higher-energy substructure in $\Lambda$ belongs to 
$d^8\underline{L}$. Then the experimental peak A is a 
$\{d^8\underline{Z},d^8\underline{L}\}$ superposition (with larger $d^8\underline{Z}$
content) and peak B results from a superposition of $d^8\underline{L}$ with part of 
the nonbonding O($2p$) spectrum. Explicit nonlocal correlations, e.g. through an
intersite Ni-Ni self-energy, appear not necessary to qualitatively account for a
sizeable Ni spectral weight within peak B. As our many-body method is (charge) self-consistent
on the lattice, {\sl implicit} features of nonlocality are included. It still may be
that {\sl explicit} nonlocal self-energies beyond DMFT enhance the Ni weight in peak B.
Thus in essence, the present study highlights the intricate entanglement between the more
basic aspects of charge-transfer physics and its highly-correlated ZR ramifications.
\begin{figure}[t]
\includegraphics*[width=8.5cm]{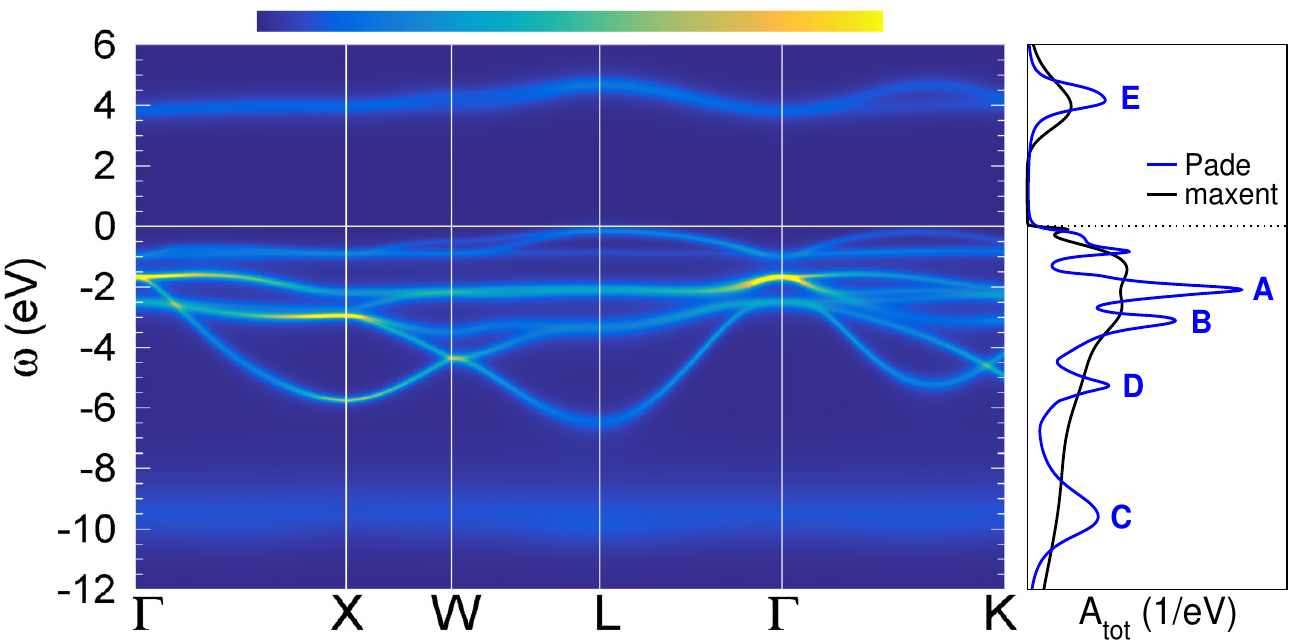}\\[-0.2cm]
\raggedright(a)\\[-0.2cm]
\begin{center}
\includegraphics*[width=7cm]{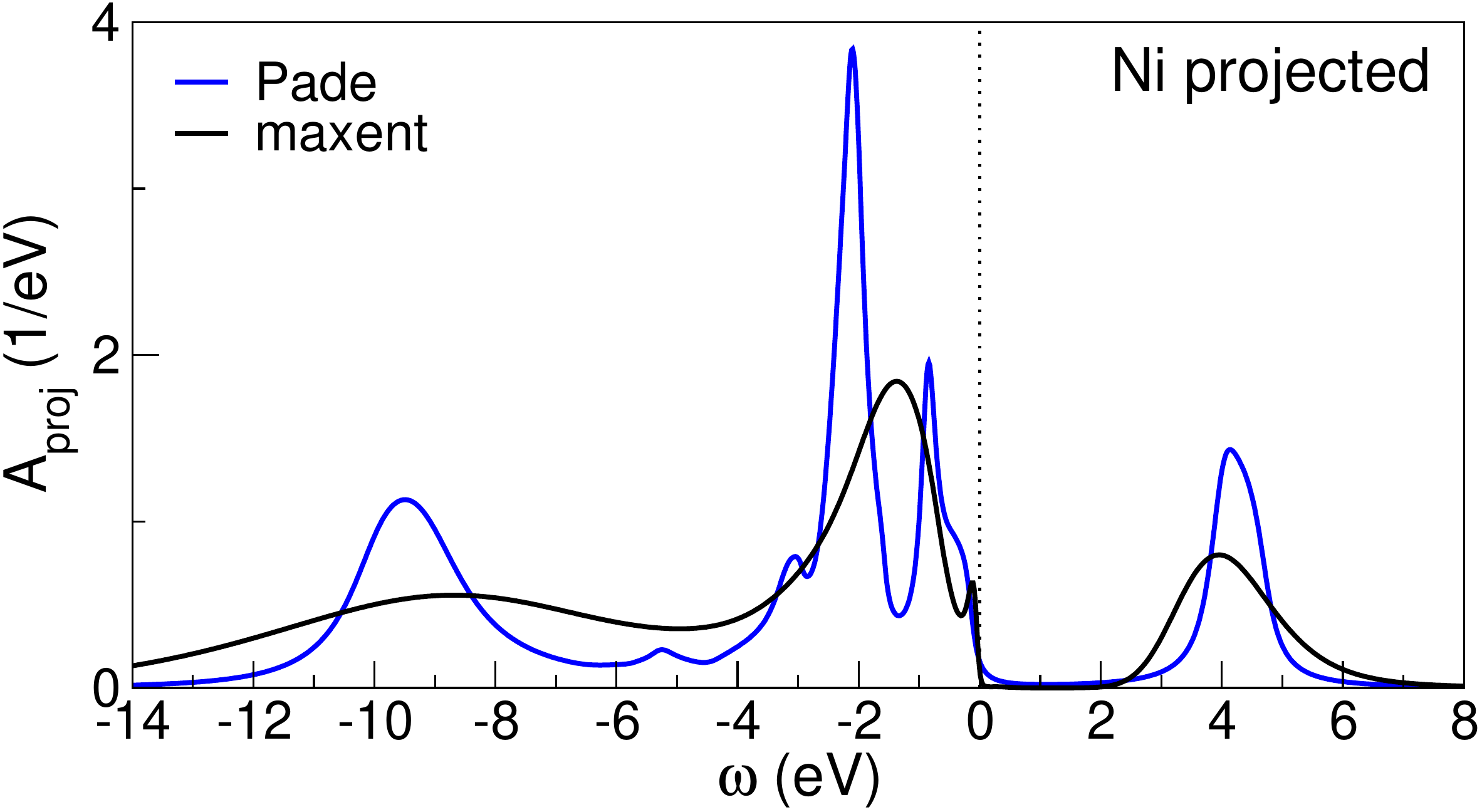}
\end{center}
\vspace*{-0.8cm}\raggedright(b)
\caption{(color online) DFT+sicDMFT {\bf k}-resolved data and comparison of different 
analytical-continuation schemes.
(a) NiO {\bf k}-resolved spectral function along high-symmetry lines (left)
and comparing total spectral function obtained from maximum-entropy method 
and Pad{\'e} method.
(b) Projected Ni$(3d)$ spectral function from Pad{\'e} method and maximum-entropy
method.
Note that the Pad{\'e} data is obtainted from analytical continuation of the Ni
self-energy, whereas the maximum-entropy data is obtained from analytical continuation
of the Bloch Green's function.}
\label{fig:kspec}
\end{figure}

Finally, we display in Fig.~\ref{fig:kspec}a the {\bf k}-resolved spectral function 
$A({\bf k},\omega)$ obtained by analytical continuation of the Matsubara self-energy via
the Pad{\'e} method. It can be seen that the upper Hubbard band still displays some dispersion.
Note that we did non include higher-energy states above the Ni-$e_g$-based bands in our 
construction of the spectral function. Additionally, Fig.~\ref{fig:kspec} provides a 
comparison between spectral data obtained via the Pad{\'e} and the maximum-entropy method.
Naturally, the former method produces sharper details. Concerning the total spectrum, all
peaks A-E are also visible with Pad{\'e}, albeit the peaks A and B are slightly shifted to
higher energies. Very close to the VBM, both analytical-continuation methods display 
spectral features in the {\bf k}-integrated spectrum. Those features are supported by the 
still visible dispersion, e.g. around the $L$ point, up to the VBM in Fig.~\ref{fig:kspec}a. 
However, this structure is not observed in the experimental 
spectrum~\cite{saw84}. This might be due to different reasons, e.g. to the suppression of 
spectral weight close to the VBM/Fermi level because of Fermi-function effects at the 
boundary of the occupied part and/or resolution issues in experiment. Novel experimental 
examinations with focus on the Ni-O hybridization in bulk NiO close to the VBM region would 
be helpful to clarify this matter. Last but not least, Fig.~\ref{fig:kspec}b shows a clearer 
projected-Ni$(3d)$ contribution to peak B at $\sim$ 3\,eV within the Pad{\'e}-based spectrum 
compared to the more-smeared maximum-entropy spectrum.

\subsubsection{Comparing with standard DFT+DMFT}
\begin{figure}[b]
\includegraphics*[width=8.5cm]{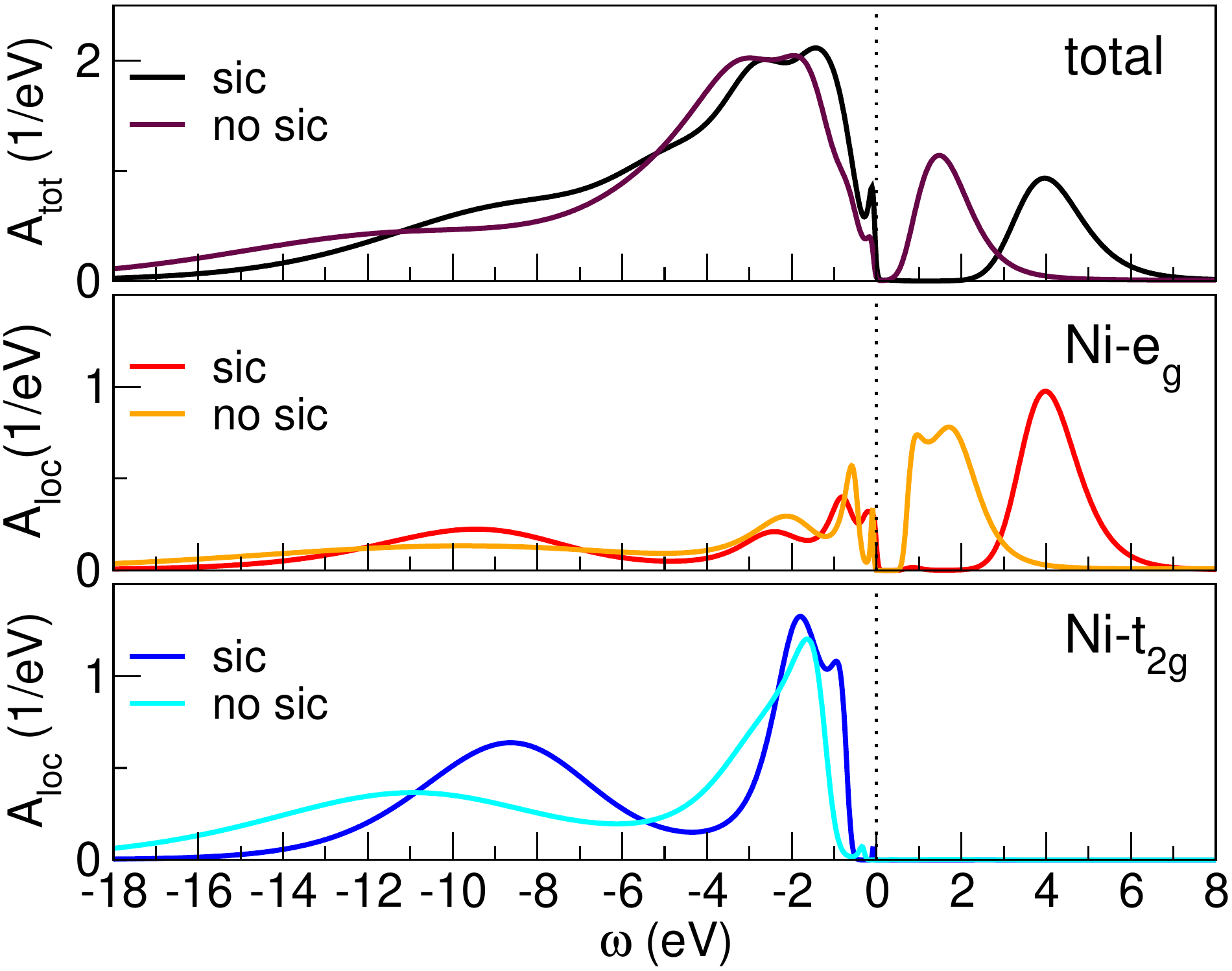}
\caption{(color online) Comparison of NiO spectral information from DFT+sicDMFT and
DFT+DMFT. Top: total spectrum, middle: local Ni-$e_g$ and bottom: local Ni-$t_{2g}$.}
\label{fig:comp}
\end{figure}
Let us now comment on the methodological aspect of our approach. In order to compare
the present scheme with the traditional DFT+DMFT method for NiO, Fig~\ref{fig:comp} shows
the total spectral function as well as the Ni-$e_g$,$t_{2g}$ local spectral function
with and without SIC for oxygen. The most obvious and striking difference concerns
the gap size: without SIC, the charge gap turns out only of order $\sim 1.5$\,eV. A similar
qualitative observation has been made by Panda {\sl et al.}~\cite{pan16}, who noted that they reached
only a small NiO charge gap with the FLL double counting in charge self-consistent DFT+DMFT.
On the occupied high-energy side, the lower Hubbard band is shifted to more negative energies
without SIC. On the other hand, the structure in the $\Lambda$ region is not dramatically
altered, albeit the fine structure appears less detailed without SIC. For instance, the
local Ni-$t_{2g}$ spectrum exhibits only a single peak in standard DFT+DMFT. Altogether,
even if focusing only on the occupied spectrum, the agreement with experiment concerning
peak signature and position is less satisfactory than with SIC. We also performed
DFT+DMFT calculations for smaller $U=8$\,eV, but the gap size did not change. 

The very fact that charge self-consistent DFT+DMFT with standard double counting fails 
in reproducing the correct paramagnetic gap size for a charge-transfer(-like) compound, is not 
that surprising. As observed, the main Mott-Hubbard physics, i.e. the formation of Hubbard
bands and their splitting in energy, is rather similar with and without SIC. However 
without SIC, the O($2p$) level is too strongly shifted in the direction of the upper 
Hubbard band, rendering the final charge gap small. Because energy-beneficial 
charge fluctuations are suppressed on the Ni site, the formalism tries to shift O($2p$)
towards the upper Hubbard band to enable as much as possible the (virtual) charge 
fluctuations between O($2p$) and Ni-$e_{g}$. Since there is no Coulomb penalty from SIC,
the charge-transfer energy shrinks. Note that there is a recent charge self-consistent
NiO calculation by Leonov {\sl et al.}~\cite{leo16}, providing a reasonable spectrum with 
only minor gap-size reduction. But in that calculation, only density-density
interactions in the local interacting Hubbard Hamiltonian have been taken into account.

Generally, one-shot DFT+DMFT 
calculations~\cite{ren06,kun07} obtain a charge gap somewhat smaller than in experiment, but
still of sensible size. This is understandable, as the O($2p$) level remains essentially
fixed in those calculations, and the DFT charge-transfer energy of $\sim 3.2$\,eV is
nearly unaltered. Hence when promoting the method to charge self-consistency, it is 
essential to include also the ligand-based Coulomb interactions. Of course, it may be 
that some other double-counting protocol can ``fix'' the problem. But the present
approach is more physical, it is applicaple to general charge-transfer problems and allows one
the use of the identical standard double-counting form for Mott-Hubbard and charge-transfer
systems without any further adjustments. Last but not least, we are confident that the correct 
interplay of Mott-Hubbard, charge-transfer and charge self-consistent processes is very 
well captured by our DFT+sicDMFT framework. 

\subsection{Li-doped NiO}
Lithium doping of NiO has first been studied in detail in the 1950s~\cite{hei57,goo58} and
remained of research significance ever since~\cite{kui89,elp92,rei95,kun07,lan07,zha18}. 
Recently, it furthermore gained interest in the context of hybrid organic-inorganic 
perovskite solar cells~\cite{jen14,che15}.
\begin{figure}[t]
\includegraphics*[width=8cm]{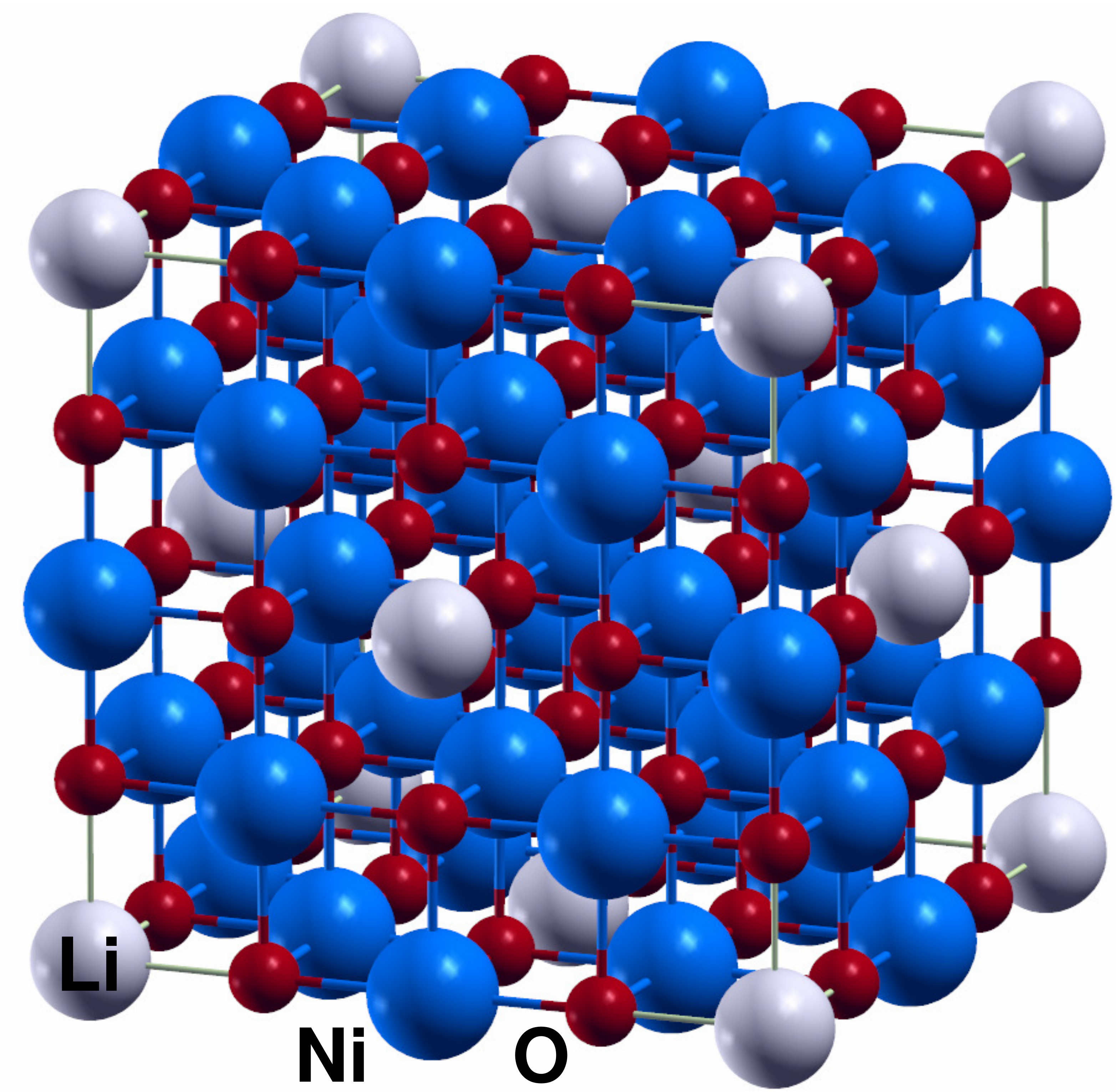}
\caption{(color online) Supercell of Li$_x$Ni$_{1-x}$O for $x=0.125$, consisting of 16
lattice sites in the primitive fcc cell. Ni: large blue, O: small red
and Li: large grey.}
\label{fig:cell}
\end{figure}

The alkali element enters as a substitutional Li$^{+}$ defect replacing Ni$^{2+}$ and thus 
providing holes to the compound. The Li$_x$Ni$_{1-x}$O system is stable for a wide $x$ 
range up to LiNiO$_2$~\cite{goo58}. Importantly, Li doping does not render NiO metallic,
but semiconducting with in-gap states appearing at $\sim 1.2$\,eV above the valence-band 
maximum~\cite{elp92,zha18}.
In the following, Li$_x$Ni$_{1-x}$O is studied for $x=0.125$ by our DFT+sicDMFT method
within a supercell approach. We reduced the stoichiometric lattice constant for the 16-atom
cell to $a_{\rm dop}=4.15\,$\AA\, in the doped case~\cite{goo58} and relaxed the atomic positions 
in antiferromagnetic LDA+U (see Fig.~\ref{fig:cell}). There are three symmetry-inequivalent 
Ni classes in the supercell. Note that structural relaxation including local Coulomb interactions 
is important, since a LDA-based relaxation leads to a too strong shift of the O sites next to Li 
towards Ni.
This results in artificial resonances in the low-energy part of the spectral function. Identical 
local Coulomb parameters as chosen for stoichiometric NiO are used for the defect problem. The 
number of Kohn-Sham projection states of is properly scaled from 8 at $x=0$ to 
59 $(=7\,{\rm Ni}\times5\,d$-${\rm orbitals}+8\,{\rm O}\times3\,p$-${\rm orbitals})$ at 
$x=0.125$.

\subsubsection{Many-body spectrum with in-gap states}
\begin{figure}[t]
(a)\hspace*{-0.4cm}\includegraphics*[width=8.5cm]{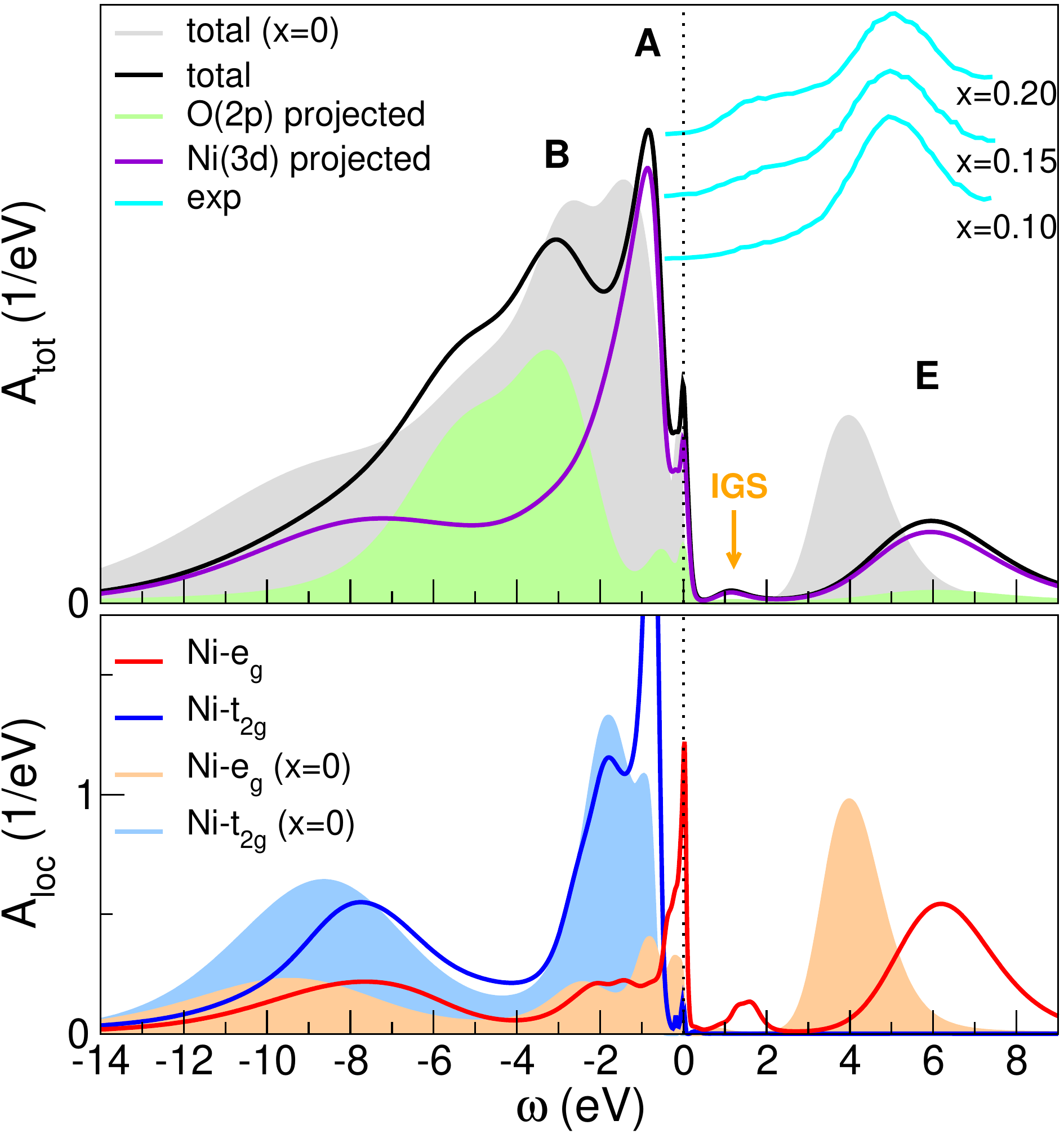}
(b)\hspace*{-0.2cm}\includegraphics*[width=8.3cm]{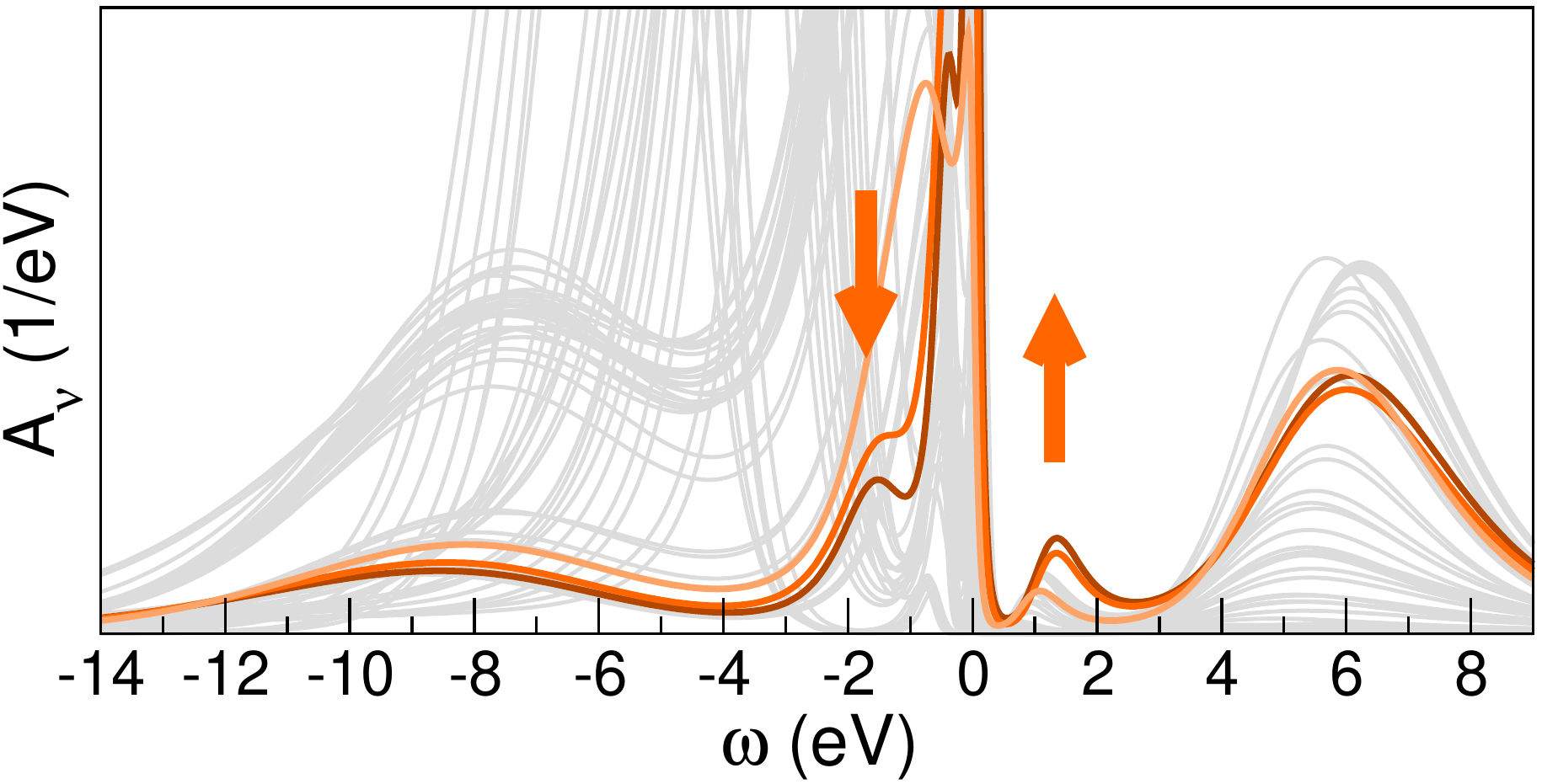}
\caption{(color online) Li$_{0.125}$Ni$_{0.875}$O spectral function from DFT+sicDMFT. 
Top: total spectrum with site- and orbital projection as well as comparison to 
stoichiometric case ($x=0$) and to inverse photoemission data~\cite{elp92} at a 
photon energy of 1486.6\,eV. In-gap states (IGS) are marked with orange arrow. 
Bottom: local Ni($3d$) spectrum with $e_g$ and $t_{2g}$ character with comparison
to stoichiometric case. (b) Bloch-resolved spectrum $A_{\nu}$ with $\nu=1\ldots 59$. 
Orange lines ordered by intensity mark four $\nu$ with decreasing contribution 
strength to the IGS. Arrows mark growth(reduction) of spectral weight.}
\label{fig:specdop}
\end{figure}
Figure~\ref{fig:specdop} depicts the collected spectral results together with 
inverse-photoemission data from van Elp {\sl et al.}~\cite{elp92}. The total spectrum in 
the upper part of Fig.~\ref{fig:specdop}a shows minor shifts and changes of intensity for 
the occupied states compared to the stoichiometric case. A shift of the spectrum to smaller 
binding energies with Li doping is also observed in experimental data~\cite{elp92,rei95,zha18}. 
Peak E in the conduction states is shifted to higher energy, which qualitatively coincides 
with available experimental data from Reinert {\sl et al.}~\cite{rei95}, yet the present theoretical
shift appears somewhat too strong. Note in addition that the inverse-photoemission 
data in Ref.~\onlinecite{elp92} were aligned at 5\,eV.
There is some minor occupation just above the VBM, i.e. the spectrum is not of perfectly strict
insulating kind. Without disorder mechanisms and/or without a local structural/electronic 
mechanism~\cite{lec18}, avoiding residual metallicity in a doped correlated insulator is very
challenging in theoretical electronic structure methods. We still believe that our spectrum
provides a reasonable account of experimentally semiconducting Li$_{0.125}$Ni$_{0.875}$O.

Concerning the joint appearance of the $d^8\underline{Z}$ and $d^8\underline{L}$ states,
the inspection of the local spectral function again turns out to be useful (see lower part of 
Fig.~\ref{fig:specdop}a). In the energy window $\Lambda=[-3.5,0]$\,eV the lowest-energy peak of 
the $t_{2g}$ spectral weight is increased and the other one of the two-peak substructure
is weakened compared to the stoichiometric case. The changes in the $e_g$ 
spectrum within this region have an even larger bearing: the stoichiometric peak at
$\sim 2.6$\,eV is nearly smeared out for $x=0.125$. This means that the vital Ni($3d$) peak 
contribution to peak B in the total spectrum is weakened. Thus the subtle balance between
$d^8\underline{Z}$ and $d^8\underline{L}$ as observed in the stoichiometric case becomes
qualitatively disturbed with Li doping. The spectral weight of $d^8\underline{Z}$ states
close to the VBM is increased.

Finally, let us focus on the in-gap states (IGS) visible in the range
$[0.7,1.7]$\,eV of the total spectrum. The energy location of the IGS is in excellent 
agreement with inverse-photoemission data~\cite{elp92,rei95,zha18}. In the local Ni 
spectrum, as already seen for the upper Hubbard band, there is no $t_{2g}$ contribution 
to the IGS. The question arises at the expense of which spectral weight do the IGS 
appear? Inspection of the weight transfers among the Bloch-resolved spectral 
functions $A_{\nu}(\omega)$ may help to answer this question. Figure~\ref{fig:specdop}b 
displays all the 59 $A_{\nu}$ functions and highlights the four ones with a major
contribution to the IGS.
From these, one observes that the 'spectral growth' of the IGS corresponds with spectral
weight at the VBM and a 'spectral shrinkage' of weight that is just somewhat deeper than 
peak A in the occupied spectrum. It is therefore tempting to directly associate 
the IGS with the weakening of the original $d^8\underline{L}$ subpeak in the local
Ni-$e_g$ spectrum as noted above.

\subsubsection{Where are the holes?}
The rigorous real-space location of the holes introduced by Li doping of NiO is a 
tenacious matter of debate~\cite{hei57,kui89,elp92,rei95,zha18}. 
Originally, data have been interpreted via the formation of Ni$^{3+}$, yet more 
recent studies favor the picture of the holes being located in the O($2p$) shell. From a 
theoretical point of view it is indeed notoriously difficult to uniquely associate 
valence charges in a condensed matter system with a specific lattice site. Depending on 
the choice of local orbitals and the kind of charge analysis, often rather different 
results are obtained. 

From a general point of view, underlined by the previous discussions of spectra, the
relevant charge-transfer physics in NiO renders the tight entanglement of Ni($3d$) and
O($2p$) in the immediate neighborhood of the VBM obvious. The Zhang-Rice doublet state
is a direct consequence thereof. Quantitatively, doping of $x$ Li ions introduces a
hole doping of $\delta=x/(1-x)$ on the remaining Ni sites. Thus in the present case of 
$x=0.125$, this amounts to $\delta=0.143$. The Ni($3d$) occupation from the local
Green's function reads $n_{d}=8.175$ at stoichiometry and $n_{d}=8.064$ with Li doping.
Hence numbers $\delta_{d}=0.111$ of effective nickel holes and 
$\delta_p=\delta-\delta_{d}=0.032$ of effective oxygen holes results here. The mixed
character of hole formation is in line with doping into the entangled Ni($3d$)-O($2p$)
VBM states.
Note that the real-space hole distribution is of course also subject to the actual choice 
of the supercell. 

\section{Summary\label{sec:sum}}
A methodological advancement of the combination of density functional theory and
dynamical mean-field theory, geared to especially address materials problems with substantial
charge-transfer character, has been presented. The combination of self-interaction 
correction on the ligand sites within the state-of-the-art charge self-consistent 
DFT+DMFT framework proves to be a powerful tool. Not only to approach long-standing 
'basic' problems of late TM oxides, but due to its efficient and easily scalable 
structure also for application to novel problems e.g. occuring from nanoscale 
structuring. There are also other theoretical approaches that may deal with electronic
correlations originating directly from ligand states. Most notably, the GW+DMFT 
scheme~\cite{bie03,cho16,pan16} has the potential to tackle such physical issues. 
But that framework is computationally much heavier than the DFT+sicDMFT scheme and 
has not yet been applied to materials with defects or dopants.

We exemplified our method for the case of stoichiometric and Li-doped NiO,
two persistently problematic correlated materials. A faithful 
description of the NiO spectrum at stoichiometry with good accordance to experimental 
findings was given. The interplay of different forms of Ni($3d$) and O($2p$)-hole states, in 
the form of $d^8\underline{Z}$ and $d^8\underline{L}$, (re)emerged~\cite{tag08} from this 
analysis. Importantly, the present scheme based on still local DMFT self-energies is
sufficient to account for the experimentally-derived hypotheses of Ni($3d$) contributions
to the double-peak structure (i.e. peak A and B) below the valence-band maximum. The
introduction of {\sl explicit} nonlocal Ni-Ni self-energies (e.g. via cluster-DMFT) is
not needed for a {\sl qualitative} appearance of such contributions. In the case of 
Li$_x$Ni$_{1-x}$O, we theoretically verified the long-standing experimental observation of 
in-gap states at $\sim 1.2$\,eV above the VBM. Our examination suggests that spectral-weight 
transfer from $d^8\underline{Z}$,$d^8\underline{L}$ into the gap region plays a relevant role in
the formation of the IGS. 

Our extended DFT+DMFT treatment of NiO does however 
not provide an end to the enduring theoretical investigation of this challenging material.
For instance, in order to improve on the classification of the competing states, 
interacting many-body and multiplet resolution on Ni, on O and inbetween, 
within a translational-invariant DMFT construction is needed. Still, the successful 
state-of-the-art DFT+DMFT approach for early/middle-row transition-metal 
oxides is thus complemented with the DFT+sicDMFT framework for late-row TM oxides.
Various problems of materials with interacting electrons, such as e.g. rare-earth nickelates or 
high-$T_{\rm c}$ cuprates, await (renewed) investigation.

\begin{acknowledgments}
We gratefully acknowledge financial support from the German Science Foundation (DFG) via 
the project LE-2446/4-1 and thank I. Leonov, L. F. J. Piper and G. A. Sawatzky 
for helpful discussions.
Computations were performed at the University of Hamburg and the JUWELS 
Cluster of the J\"ulich Supercomputing Centre (JSC) under project number hhh08.
\end{acknowledgments}

\bibliographystyle{apsrev}
\bibliography{bibextra}

\end{document}